\documentclass[preprint,prl,showpacs,floatfix,superscriptaddress]{revtex4-1}%
\usepackage{graphicx}
\usepackage{dcolumn}
\usepackage{bm}
\usepackage{amssymb}
\usepackage{amsmath}
\usepackage{amsfonts}
\usepackage{float}
\begin{document}
%\begin{flushleft}
%\texttt{submitted to Physical Review Letters}		
%\end{flushleft}

\title{Imaging the He$_{2}$ quantum halo state using a free electron laser}

\author{S. Zeller*}
\author{M. Kunitski}
\author{J. Voigtsberger}
\author{A. Kalinin}
\author{A. Schottelius}
\author{C. Schober}
\author{M. Waitz}
\author{H. Sann}
\author{A. Hartung}
\author{T. Bauer}
\author{M. Pitzer}
\author{F. Trinter}
\author{C. Goihl}
\author{C. Janke}
\author{M. Richter}
\author{G. Kastirke}
\author{M. Weller}
\author{A. Czasch}
\affiliation{Institut f\"ur Kernphysik, Goethe-Universit\"at Frankfurt, 60438
Frankfurt am Main, Germany}
\author{M. Kitzler}
\affiliation{Photonics Institute, Vienna University of Technology, Gu\ss hausstra\ss e 27, 1040 Vienna, Austria}
\author{M. Braune}
\affiliation{Deutsches Elektronen-Synchrotron DESY, Notkestra\ss e 85, 22607 Hamburg, Germany}
\author{R. E. Grisenti}
\affiliation{Institut f\"ur Kernphysik, Goethe-Universit\"at Frankfurt, 60438
Frankfurt am Main, Germany}
\affiliation{GSI Helmholtz Centre for Heavy Ion Research, Planckstra\ss e 1, 64291 Darmstadt, Germany}
\author{W. Sch\"ollkopf}
\affiliation{Department of Molecular Physics, Fritz-Haber-Institut, Faradayweg 4-6, 14195 Berlin, Germany}
\author{L. Ph. H. Schmidt}
\author{M. Sch\"offler}
\affiliation{Institut f\"ur Kernphysik, Goethe-Universit\"at Frankfurt, 60438
Frankfurt am Main, Germany}
\author{J. B. Williams}
\affiliation{Department of Physics, University of Nevada, 1664 N. Virginia Street, Reno, NV 89557, USA}
\author{T. Jahnke}
\author{R. D\"orner*}
\affiliation{Institut f\"ur Kernphysik, Goethe-Universit\"at Frankfurt, 60438
Frankfurt am Main, Germany}

\date{\today}

\begin{abstract}
We report on Coulomb explosion imaging of the wavefunction of the quantum halo system He$_2$.
Each atom of this system is ionized by tunnel ionization in a femto second laser pulse and in a second experiment by single photon ionization employing a free electron laser. We visualize the exponential decay of the probability density of the tunneling particle over distance for over two orders of magnitude up to an internuclear distance of 250~\AA. By fitting the slope of the density in the tunneling regime we obtain a binding energy of 151.9~$\pm$~13.3~neV, which is in agreement with most recent calculations \cite{Przybytek2010}.
\end{abstract}

\maketitle

\begin{figure}[tb]
\centering
\includegraphics[width=0.6\columnwidth]{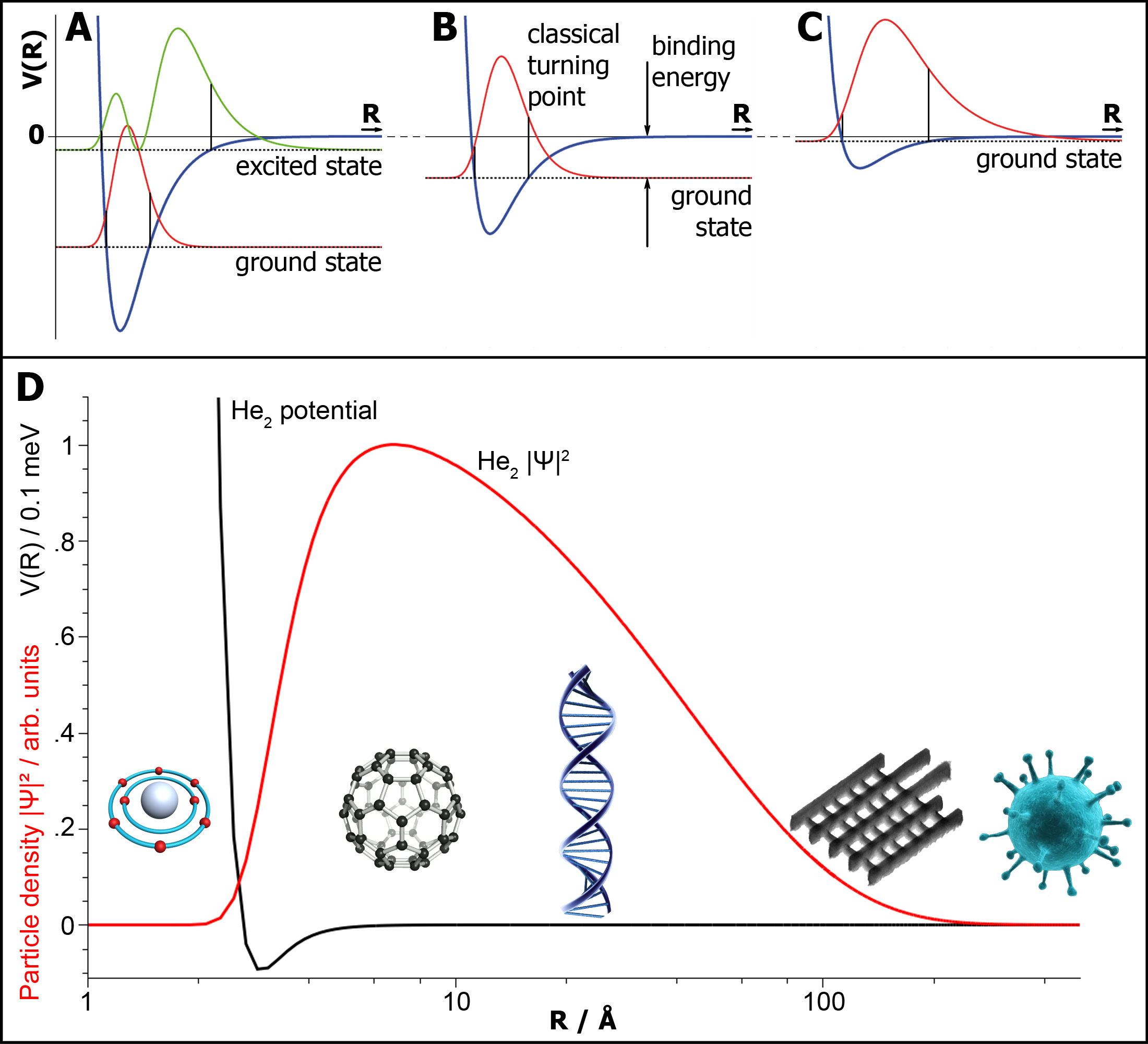}
\caption{A shallow short range potential holding a ground and an excited state (A). As the potential depth decreases (B) the excited state becomes unbound, leaving only the ground state. Further decrease (C) leads to the particle probability density distribution leaking more into the classically forbidden region. In the extreme case of the helium dimer (D) (note the logarithmic R-scale) this effect allows the wavefunction to extend to sizes of fullerenes, the diameter of DNA and even small viruses (He$_2$ potential and wavefunction taken from \cite{Przybytek2010}): while the classical turning point is located at 13.6~\AA~the overall wavefunction extends to more than 200~\AA.}
\label{fig_theory}
\end{figure}

\begin{figure}[tb]
\centering
\includegraphics[width=0.6\columnwidth]{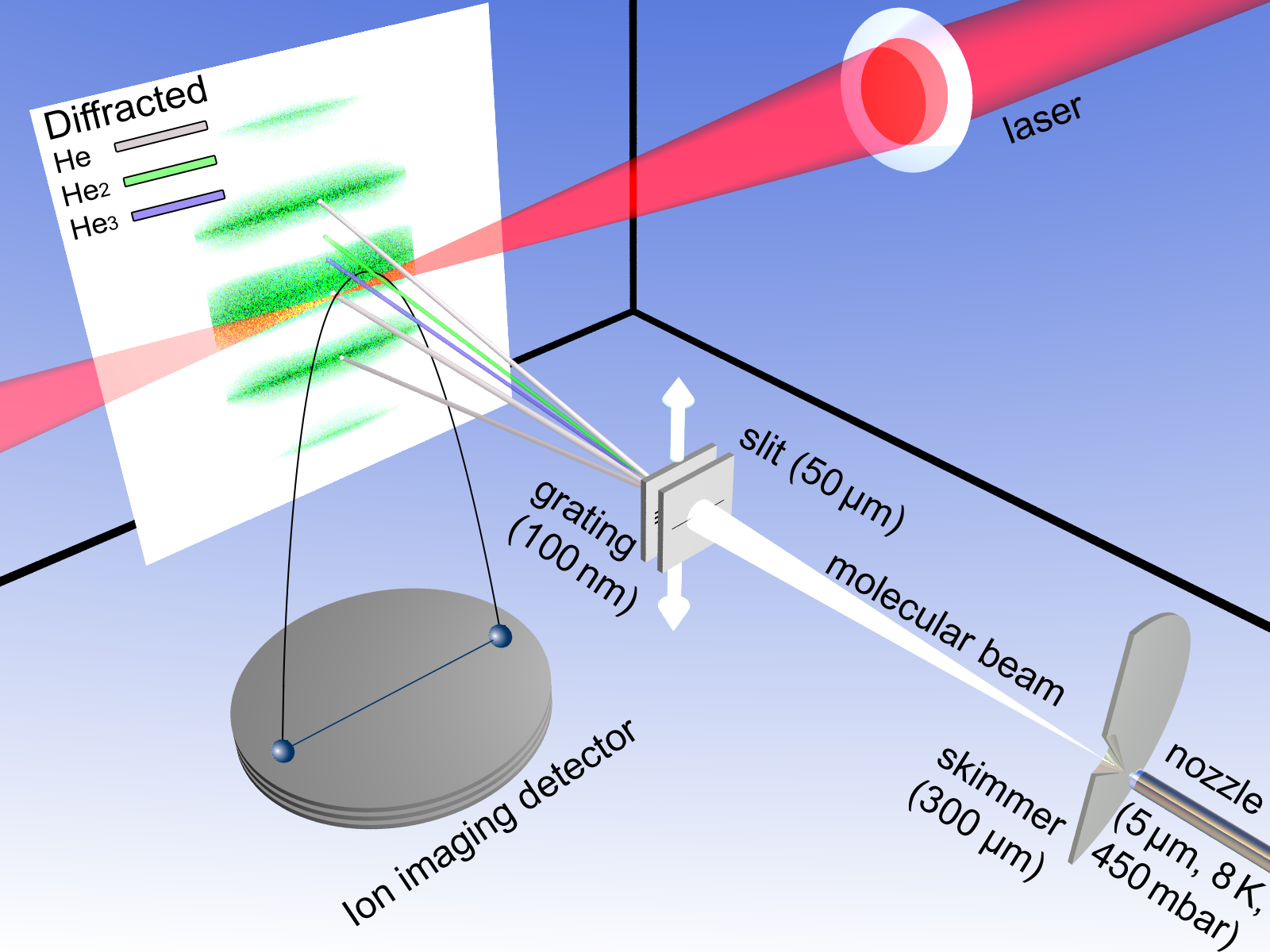}
\caption{Overlap between laser focus and a pure helium dimer beam, created by a molecular beam diffracted at a nanograting. Distances between the beam elements were as follows: nozzle to skimmer 14~mm, skimmer to slit 332~mm, slit to grating 30~mm, grating to focus 491~mm. The focus diameter was about 20~$\mu$m.}
\label{fig_setup}
\end{figure}

\begin{figure*}[tb]
\centering
\includegraphics[width=1.0\columnwidth]{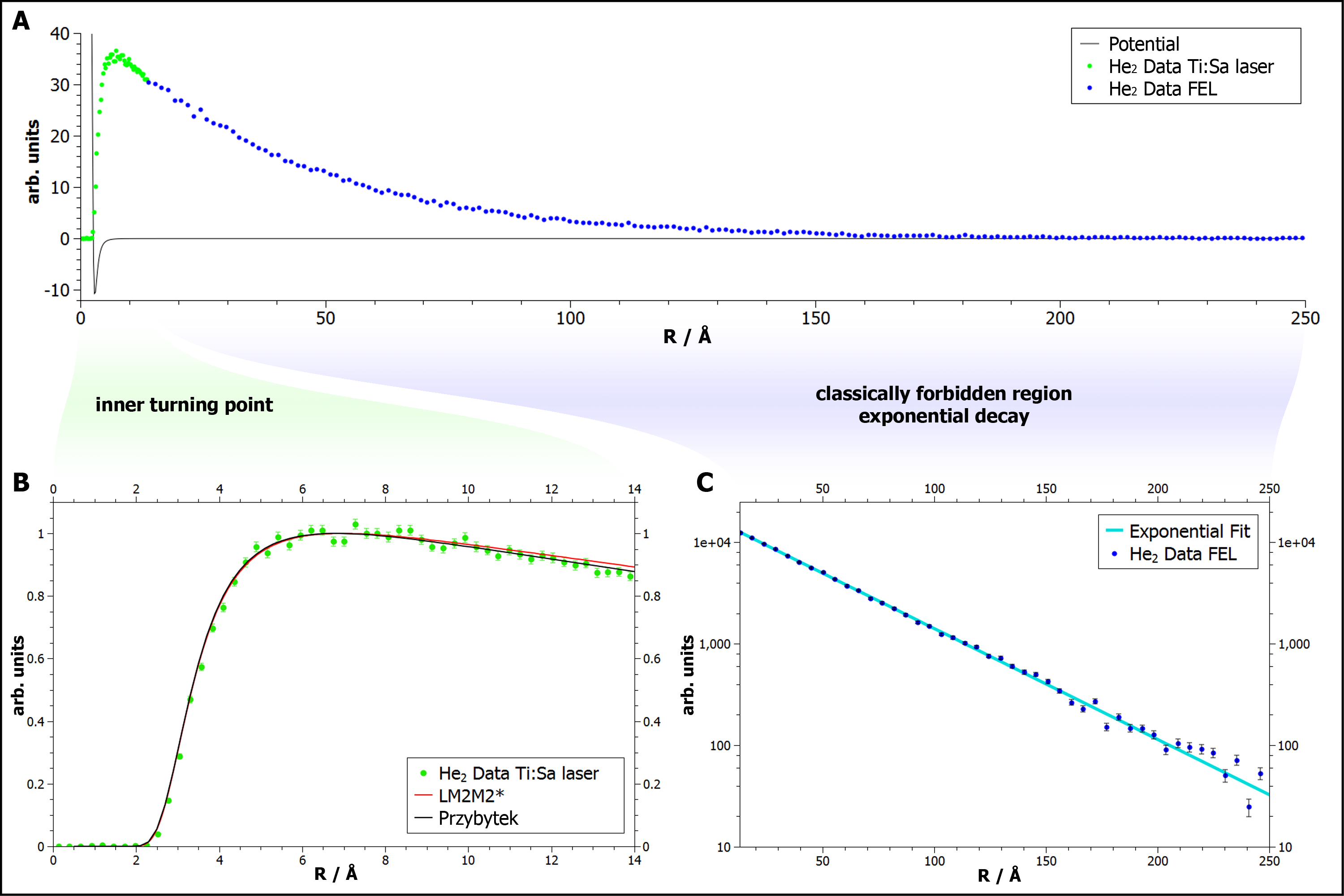}
\caption{Measurement of the helium dimer wavefunction (A). Two detailed views show the important features of this quantum system: The region of the inner turning point (B) is in agreement with theoretical predictions LM2M2*\cite{LuoInfluence1993} and Przybytek\cite{Przybytek2010}, and the exponential decay in the classical forbidden region (C). A helium dimer binding energy of 151.9~$\pm$~13.3~neV is obtained from the exponential slope. The electron recoil has to be taken into account to conclude from the slope shown in C to the value of the binding energy (see text for details).}
\label{fig_experiment}
\end{figure*}

\begin{figure}[tb]
\centering
\includegraphics[width=0.6\columnwidth]{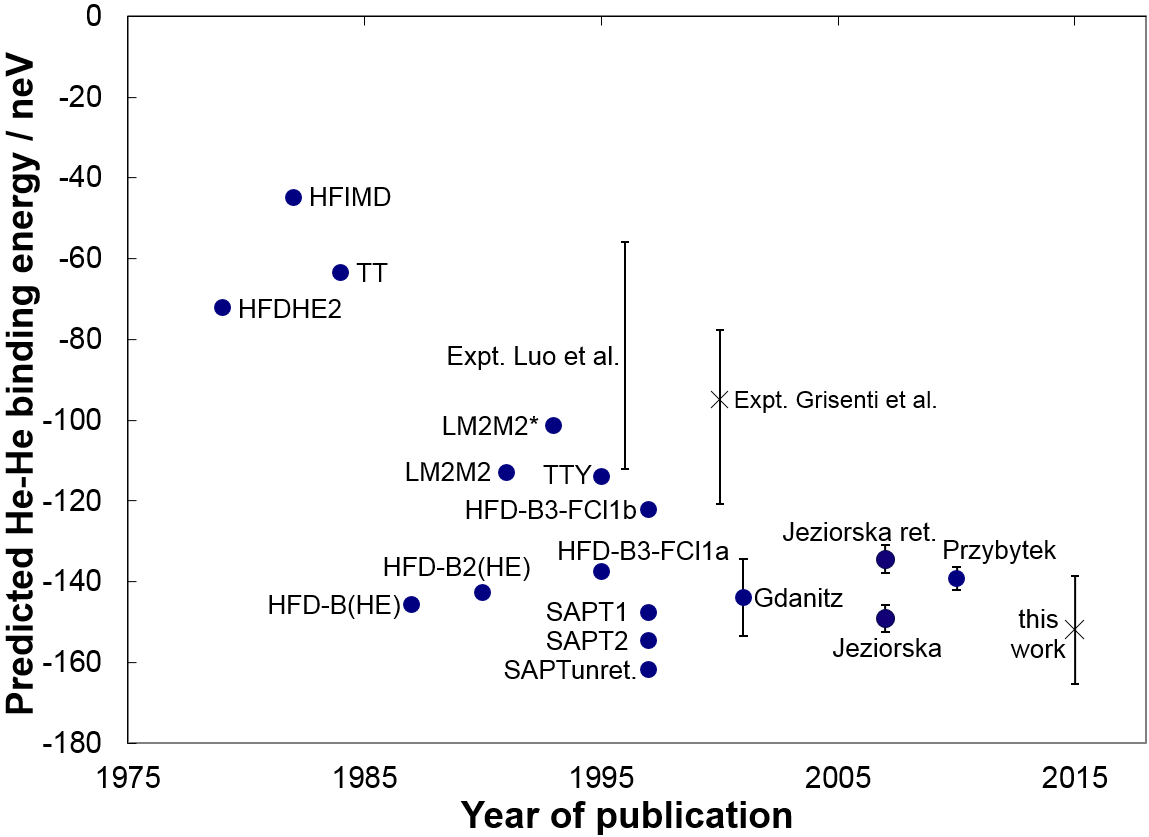}
\caption{The predicted values for the helium dimer binding energy using various theoretical calculations (HFDHE2\cite{Aziz1979}, HFIMD\cite{Feltgen1982}, TT\cite{Tang1984}, HFD-B(HE)\cite{Aziz1987}, HFD-B2\cite{Aziz1990}, LM2M2\cite{Aziz1991}, LM2M2*\cite{LuoInfluence1993}, TTY\cite{Tang1995}, HFD-B3-FCl1a\cite{Aziz1995}, HFD-B3-FCl1b, SAPT\cite{Janzen1997,Korona1997}, Gdanitz\cite{Gdanitz2001}, Jeziorska, Jeziorska ret.\cite{Jeziorska2007} and Przybytek\cite{Przybytek2010}) are displayed alongside experimental measurements from Luo et al.\cite{Luo1996}, Grisenti et al.\cite{Grisenti2000} and the present work.}
\label{fig_vergleich}
\end{figure}

%\section{Introduction}
Quantum tunneling is a ubiquitous phenomenon in nature and crucial for many technological applications. It allows quantum particles to reach regions in space which are energetically not accessible according to classical mechanics. In this tunneling region the particle density is known to decay exponentially. This behavior is universal across all energy scales from MeV in nuclear physics, to eV in molecules and solids, and to neV in optical lattices. For bound matter the fraction of the probability density distribution in this classically forbidden region is usually small. For shallow short range potentials this can change dramatically as shown in Fig.~\ref{fig_theory}: upon decreasing the potential depth excited states are expelled one after the other as they become unbound (transition from Fig.~\ref{fig_theory}~A to B). A further decrease of the potential depth effects the ground state as well, as more and more of its wavefunction expands into the tunneling region (Fig.~\ref{fig_theory}~C/D). Consequently, at the threshold (i.e. in the limit of vanishing binding energy) the size of the quantum system expands to infinity. For short range potentials this expansion is accompanied by the fact that the system becomes less “classical” and more quantum-like. Systems existing near that threshold (and therefore being dominated by the tunneling part of their wavefunction) are called “quantum halo states” \cite{Riisager1994}. These are, for example, known from nuclear physics where $^{11}$Be and $^{11}$Li form halo states \cite{Jensen2004,Hansen2003,Tanihata1996}.\\
One of the most extreme examples of such a quantum halo state can be found in the realm of atomic physics: the helium dimer (He$_2$). It is bound by the van der Waals force only and the He-He interaction potential (see Fig. 1D) has a minimum of about 1~meV at an internuclear distance of about 3~\AA~(0.947~meV / 2.96 \AA \cite{Przybytek2010}). For a long time it was controversial whether already the zero point energy of the helium dimer is larger than the depth of the potential well and thus whether the helium dimer exists as a stable molecule at all. While $^{3}$He$^{4}$He is indeed unbound because of its bigger zero point energy, stable $^{4}$He$_2$ was finally found experimentally in 1993/94 \cite{Schoellkopf1994,Luo1993}. It turns out, that He$_2$ has no bound excited rotational states as already the centrifugal force associated with 1$\hbar$ of angular momentum leads to dissociation. Experiments using matter wave diffraction confirmed the halo character of He$_2$ by measuring a mean value of the internuclear distance of 52~\AA \cite{Grisenti2000}. This is in agreement with some theoretical predictions, but in conflict with the most recent calculations\cite{Przybytek2010}. Resolving this conflict is of importance also for the planned redefinition of the Kelvin, unit of thermodynamic temperature, in terms of the Boltzmann constant\cite{Lin2013}. Thermometry today uses theoretical values for the thermal conductivity and viscosity of helium. Those properties are based on the same He-He interaction potentials used to calculate the He$_2$ binding energy, which was shown to be incompatible with previous experiments\cite{Grisenti2000,Luo1996} (see \cite{Cencek2012} for a more detailed discussion).\\
At the same time its quantum halo character makes He$_2$ a prime candidate for visualizing the predicted universal exponential decrease of a tunneling wavefunction in an experiment by triggering a Coulomb explosion with a free electron laser (FEL). Coulomb explosion imaging is a well-established technique first employed in ion beam experiments\cite{Kanter1989}. For chiral molecules fragmented by femtosecond laser pulses it has been successfully used to identify enantiomers\cite{Pitzer2013}. For diatomics it has been shown to reveal subtle details of the wavefunction at the quantum limit of position measurements\cite{Schmidt2012}. Most recently we have used the technique to study the structure of He$_3$\cite{Voigtsberger2014} and to discover the Efimov state of He$_3$\cite{Kunitski2015}. In the two latter studies we have combined Coulomb explosion imaging with cluster mass selection by matter wave diffraction\cite{Schoellkopf1994}. The COLTRIMS reaction microscope used in \cite{Voigtsberger2014,Kunitski2015} was the same as used in the present study. In the present study we employ for the first time single photon ionization by FEL radiation in stead of sequential tunnel ionization by an 800~nm laser pulse. Only this use of single photon ionization allows for the precise determination of the slope of the exponential decay of the wavefunction we achieve in the current measurement.
\\
In the corresponding experiment presented here, helium clusters were produced by expanding helium gas through a 5~$\mu$m nozzle. It was cooled down to 8~K and a driving pressure of 450~mbar was applied, which maximizes the dimer content in the molecular beam\cite{Kunitski2015}. To obtain a pure helium dimer target beam we made use of matter wave diffraction \cite{Schoellkopf1994}. All clusters have the same velocity but can be sorted by mass as their diffraction angle behind a transmission grating (100~nm period) depends on their de Broglie wavelengths ($\lambda =$ h/mv, with Planck’s constant~h, mass~m and velocity~v). That way only dimers reach the interaction region while the dominant fraction of atomic helium as well as the share of helium trimers present at the chosen gas expansion conditions get deflected away from the ionization region. Figure \ref{fig_setup} shows a schematic of the setup.\\
In two experimental campaigns both atoms of the dimer were singly ionized employing either single photon ionization using photons provided by a free electron laser (FLASH, $<$100~fs, 18.5~nm) or tunnel ionization using a strong ultrashort laser field (Ti:Sa laser, Dragon KMLabs, 780~nm). The two positively charged ions repel each other, resulting in a Coulomb explosion. The ionic momenta acquired in this explosion were measured by cold target recoil ion momentum spectroscopy (COLTRIMS)\cite{Jagutzki2002,Ullrich2003,Jahnke2004}. A homogeneous electric field of 4.41~V/cm (at FEL) / 3.09~V/cm (at Ti:Sa laser) guides the ions to the detector. It measures time-of-flight and position of impact using micro channel plates (MCP) and delay line anodes\cite{Jagutzki2002}. In the FEL radiation as well as in pulses of 800~nm photons the ionization of the two atoms occurs fast compared to the nuclear motion, thus triggering an instantaneous Coulomb explosion of the repelling ionized particles. The Coulomb explosion converts the potential energy of the two ions located at an internuclear distance R into a released kinetic energy (KER) according to
\begin{equation}
R = \frac{1}{KER}.
\label{eq_1}
\end{equation}  
By recording a large number of Coulomb explosion events a distribution of measured distances R (as shown in Fig. \ref{fig_experiment}A) is obtained. It represents a direct measurement of the square of the helium dimer wavefunction~\textbar$\Psi$\textbar$^2$. The classically allowed part of~\textbar$\Psi$\textbar$^2$ provides a cross-check for our measurement as it falls off steeply at the inner turning point of the helium dimer potential and theoretical calculations agree well on the location of the turning point. A comparison of our measured probability density distribution close to the inner turning point and some theoretical predictions are shown in Fig. \ref{fig_experiment}B. Here two exemplary theoretical curves\cite{Przybytek2010,LuoInfluence1993} are depicted along with a measurement conducted at our Ti:Sa laser as it provides very high resolution and statistics for small internuclear distances.\\
The classically forbidden part of~\textbar$\Psi$\textbar$^2$ is shown in Fig. \ref{fig_experiment}C on a logarithmic scale. For internuclear distances larger than 30~\AA~the helium dimer potential is two orders of magnitude smaller than the predicted ground state binding energy and thus can safely be approximated to zero. Accordingly, the wavefunction is approximated in this region by the solution of the Schr\"odinger equation below a steplike barrier, which is given by
\begin{equation}
\Psi(R) \propto e^{-\sqrt{\frac{2m}{\hbar^2}E_{bind}R}}.
\label{eq_2}
\end{equation}  
As the mass m and Planck’s constant $\hbar$ are fixed, the only variable defining the slope of the exponential decay is the binding energy E$_{bind}$. Therefore the binding energy can be extracted from the measurement by an exponential fit to the pair-distance distribution in the region between 50~a.u. and 300~a.u., as depicted in Fig. \ref{fig_experiment}C. From the fit we obtain a helium dimer binding energy of 151.9~$\pm$~13.3~neV, after accounting for the electron recoil as outlined below.\\
The theoretical value for the binding energy was under dispute for many years \cite{Aziz1979,Feltgen1982,Tang1995,Gdanitz2001}. Predictions range from 44.8~neV\cite{Feltgen1982} to 161.7~neV\cite{Janzen1997}. Recently calculations became available which include quantum electrodynamical effect, relativistic effects and go beyond the Born Oppenheimer approximation. These supposedly most precise calculations predict a binding energy of 139.2~$\pm$~2.9~neV\cite{Przybytek2010}, which is in disagreement with the most recent experimental value of 94.8~+25.9/-17.2~neV obtained in pioneering experiments by evaluating matterwave diffraction patterns and relying on a detailed theoretical modelling of the interaction of the  dimer with the grating surface\cite{Grisenti2000}. The present value of 151.9~$\pm$~13.3~neV is in good agreement with the prediction of Przybytek et al.\cite{Przybytek2010} (139.2~$\pm$~2.9~neV) and in clear disagreement with the predictions from some He-He interaction potentials, including the popular TTY\cite{Tang1995} and LM2M2\cite{Aziz1991} potentials yielding 114 and 113~neV, respectively. Figure~\ref{fig_vergleich} displays the evolution of theoretical predictions over the years.\\
Before concluding we add a discussion of the effects which contribute to the error of $\pm$13.3~neV which we give for our value of the binding energy. The major contributions to this error are the calibration of the COLTRIMS machine and the deviations from the axial recoil approximation for the Coulomb explosion.
In the COLTRIMS spectrometer the crucial parameters are the absolute value of the electric field in the spectrometer and the position calibration of the detector.\\
The electric field was obtained by measuring the kinetic energy release spectrum of the N$_2$ breakup which provides very narrow peaks. Transitions from D$^{3}\Pi$g and D$^{1}\Sigma$u$^{+}$ into continuum could be identified and met reference measurements\cite{Lundqvist1996} with a mean relative deviation of 0.054\%. This yielded the calibration of the momentum component along the time-of-flight direction of the spectrometer.\\
The position calibration was done by comparing the momentum component in the time-of-flight direction with the ones perpendicular to it. For this purpose we performed two calibration measurements with isotropic dissociation channels (N$_2$O / Ne$_2$). Most relevant, due to energetic proximity to the helium dimer breakup, is the N$_2$O channel at 0.16~eV KER with a mean relative deviation of 6.2\%, while additional channels yield a smaller deviation with 0.62\% (N$_2$O at 0.36~eV) and 0.15\% (Ne$_2$ at 4.4~eV).\\
For the experiment at FLASH, despite excellent vacuum conditions (8$\cdot$10$^{-12}$~mbar), an average of about 50 ions were collected for every FEL pulse. The majority of ions were charged hydrogen atoms or molecules with short times-of-flight, which could be gated out by software during data acquisition prior to writing to the hard drive. Nevertheless the MCP endured constant stress which led to a drop in detection efficiency in the center of the detector. The detection efficiency was corrected to its normal level using a residual gas calibration measurement with a Gaussian shaped correction function containing a 5.5\% uncertainty. This leads to $\pm$1~neV uncertainty on the binding energy. In addition random coincidences from ionizations of two independent helium ions from the residual gas were subtracted. The error resulting from this background subtraction is small in comparison to errors discussed above ($\pm$0.4~neV). We also have excluded breakups recorded in the detector plane (with a tolerance of $\pm$33.5$^{\circ}$) as indistinguishable background and potentially deadtime effects compromised the data here.\\
To image the exact shape of the probability density distribution by Coulomb explosion imaging the ionization probability has to be independent of the internuclear distance. Two consecutive tunnel ionization steps can be influenced by enhanced ionization\cite{Wu2013}, an effect which depends on the internuclear distance. The steep rise of the probability density at the inner turning point is not very sensitive to this effect and could consequently be imaged by our experiment with an 800~nm laser pulse, which has superior statistics compared to the FEL experiment (see Fig.~\ref{fig_experiment}). For the exponential region of the probability density we aim for a high precision determination of the slope. We therefore used photons from the free electron laser FLASH to ionize both atoms of the dimer by single photon absorption. Compared to an 800~nm laser pulse this has the additional advantage that the electron energy, and thus the recoil of the electrons onto the nuclei, is much better controlled and has an upper threshold.\\
The initial ion energy during the Coulomb breakup has to be either zero or well defined as equation~\ref{eq_1} assumes that the KER only results from the potential energy between the two point charges and that there is no additional energy from other sources. The two most important sources of such additional energy are the zero point kinetic energy from the bound state \cite{Schmidt2012} before ionization and the energy transferred during the ionization process by recoil of the escaping electron.\\
The first is negligible for He$_2$, because the depth of the potential well is only 1~meV. We have also confirmed that by calculating the Coulomb explosion quantum mechanically. We found no difference in the KER between the classical calculation using equation~\ref{eq_1} and the quantum calculation which automatically includes the initial state zero point motion (see\cite{Schmidt2012}).\\
The energy transferred to the two nuclei during the ionization process at the FEL is given by the recoil of the two electrons. The sum momentum distribution of two electrons with a kinetic energy of E$_{\gamma}$ -- I$_P$~=~42.4~eV each was calculated and is reflected in the measured data. For two independent ionization events the distribution of the sum momenta and the momentum difference of the electrons are equal. While the sum momentum cancels out in the KER calculation the relative momentum adds to it and increases the measured KER. This reduces the slope of the exponential decaying function by 12.1~neV. Taking this into account we obtain a binding energy value of 151.9~neV $\pm$1.7(stat) $\pm$10.2(calib) $\pm$1.4(corr)~neV from our experiment. The statistical error is the error of the fit caused by the statistics of the data points, the calibration error is the uncertainty of the calibration of our COLTRIMS reaction microscope as discussed above and the error labeled (corr) is the estimated error on the correction procedure compensating the detector efficiency and subtraction of random coincidences.\\
In conclusion the helium dimer is a remarkable example of a system existing predominantly in the quantum mechanical tunneling regime. We were able to reveal the full shape of the wavefunction experimentally. The measured data confirms the universal exponential behavior of wavefunctions under a potential barrier on unprecedented scales and yields a revised experimental value for the binding energy of the helium dimer, which has been under dispute for more than 20~years.

\acknowledgments  The experimental work was supported by a Reinhart Koselleck project of the Deutsche Forschungsgemeinschaft. We are grateful for excellent support by the staff of FLASH during our beamtime. We thank R. Gentry and M. Przybytek for providing their theoretical results in numerical form. 

%\bibliographystyle{apsrev4-1}
%\bibliographystyle{unsrt}
%\bibliographystyle{ieeetr}
%\bibliography{literatur_He2_Zeller} %Eine Datei 'literatur_He2_Zeller.bib' wird hierfür benötigt.


\end{document}